\documentclass[acmsmall,screen,nonacm]{acmart}

\usepackage{prelude}

\acmJournal{PACMPL}

\setcopyright{none}

\usepackage{booktabs}   
\usepackage{subcaption} 

\begin{document}

\title{Debugging Trait Errors as Logic Programs}


\author{Gavin Gray}
\orcid{0000-0002-2960-1198}
\affiliation{
  \department{Department of Computer Science}
  \institution{ETH Z\"{u}rich}
  \city{Z\"{u}rich}
  \country{Switzerland}
}
\email{gavin.gray@inf.ethz.ch}

\author{Will Crichton}
\orcid{0000-0001-8639-6541}
\affiliation{
  \department{Department of Computer Science}
  \institution{Brown University}           
  \city{Providence}
  \state{Rhode Island}
  \postcode{02912}
  \country{USA}                    
}
\email{wcrichto@brown.edu}

\begin{abstract}
Rust uses traits to define units of shared behavior. Trait constraints build up an implicit set of first-order hereditary Harrop clauses which is executed by a powerful logic programming engine in the trait system. But that power comes at a cost: the number of traits in Rust libraries is increasing, which puts a growing burden on the trait system to help programmers diagnose errors. Beyond a certain size of trait constraints, compiler diagnostics fall off the edge of a complexity cliff, leading to useless error messages. Crate maintainers have created ad-hoc solutions to diagnose common domain-specific errors, but the problem of diagnosing trait errors in general is still open. We propose a trait debugger as a means of getting developers the information necessary to diagnose trait errors in any domain and at any scale. Our proposed tool will extract proof trees from the trait solver, and it will interactively visualize these proof trees to facilitate debugging of trait errors.
\end{abstract}




\maketitle

\section{Introduction}

Rust is a systems programming language that provides strong memory safety guarantees through ownership. A less-touted but equally-important aspect of Rust's design is its trait system. Similar to typeclasses in Haskell, traits in Rust define units of shared behavior which can bound generic types. For example, this snippet illustrates a \rss{ToString} trait for converting values to strings:

\begin{minted}{rust}
// A trait definition establishes the unit of shared behavior.
trait ToString {
  fn to_string(&self) -> String;
}

// A trait implementation associates a type with a trait. Logically, this is the fact:
//   (i32, i32): ToString.
impl ToString for (i32, i32) {
  fn to_string(&self) -> String {
    format!("({}, {})", self.0, self.1)
  }
}

// An implementation can be parameteric. Logically, this is the rule:
//   Vec<T>: ToString :- T: ToString.
impl<T: ToString> ToString for Vec<T> {
  fn to_string(&self) -> String {
    let s = self.iter().map(|v| v.to_string()).collect::<Vec<_>>().join(", ");
    format!("[{s}]")
  }
}

// A trait method is normally invoked with the dot operator. Logically, this is the query:
//   ?- Vec<(i32, i32)>: ToString
fn main() {
  let v = vec![(0, 1), (2, 3)];
  println!("{}", v.to_string());
}
\end{minted}

To call a trait method like \rss{v.to_string()}, the Rust compiler must determine that the type of \rss{v} satisfies the conditions required to call \rss{.to_string()}. As suggested by the Prolog-esque syntax in the comments above, this problem reduces to logic programming. A trait is a predicate, a non-parameterized implementation is a fact, a parameterized implementation is a rule, and a required trait bound is a query. This analogy is made explicit by Chalk  \citegithub{https://github.com/rust-lang/chalk}, an implementation of Rust's trait solver within a generic logic programming framework.

If trait solving is logic programming, then debugging trait errors is debugging logic programs. Considering the current popularity of Prolog, the Rust compiler goes to great lengths to obscure this connection. In fact, the compiler does a heroic amount of work to help its users debug trait errors. For instance, if one tries to call \rss{.to_string()} on a \rss{Vec<i32>}, then Rust gives a handy diagnostic that localizes the root cause to the vector's type parameter:

\begin{minted}{text}
error[E0599]: the method `to_string` exists for struct `Vec<i32>`, 
              but its trait bounds were not satisfied
   --> src/main.rs:105:20
    |
105 |   println!("{}", v.to_string());
    |                    ^^^^^^^^^ method cannot be called on `Vec<i32>` 
    |                              due to unsatisfied trait bounds
    |
note: trait bound `i32: ToString` was not satisfied
   --> src/main.rs:93:9
    |
93  | impl<T: ToString> ToString for Vec<T> {
    |         ^^^^^^^^  --------     ------
    |         |
    |         unsatisfied trait bound introduced here
\end{minted}

Essentially, the compiler contains a collection of heuristics that dictate when to provide a specific kind of diagnostic. These heuristics have served Rust programmers well through the early years of the language. The impetus for our work is that the nature of Rust programs is changing. Modern Rust libraries have many more traits than before. Rust programmers are constructing precariously large (and entirely invisible!) Prolog programs. When the trait solver fails, the compiler is increasingly incapable of being the omniscient Prolog debugger that always points to the root cause. We believe that Rust developers are in need of a ``trait debugger,'' a more powerful tool for diagnosing trait errors. The goal of this paper is to explore: why is a trait debugger necessary? And how might we build one?

\section{When Traits Go Wrong}

\begin{figure}[hbtp]
    \centering
    \includegraphics[width=0.85\linewidth]{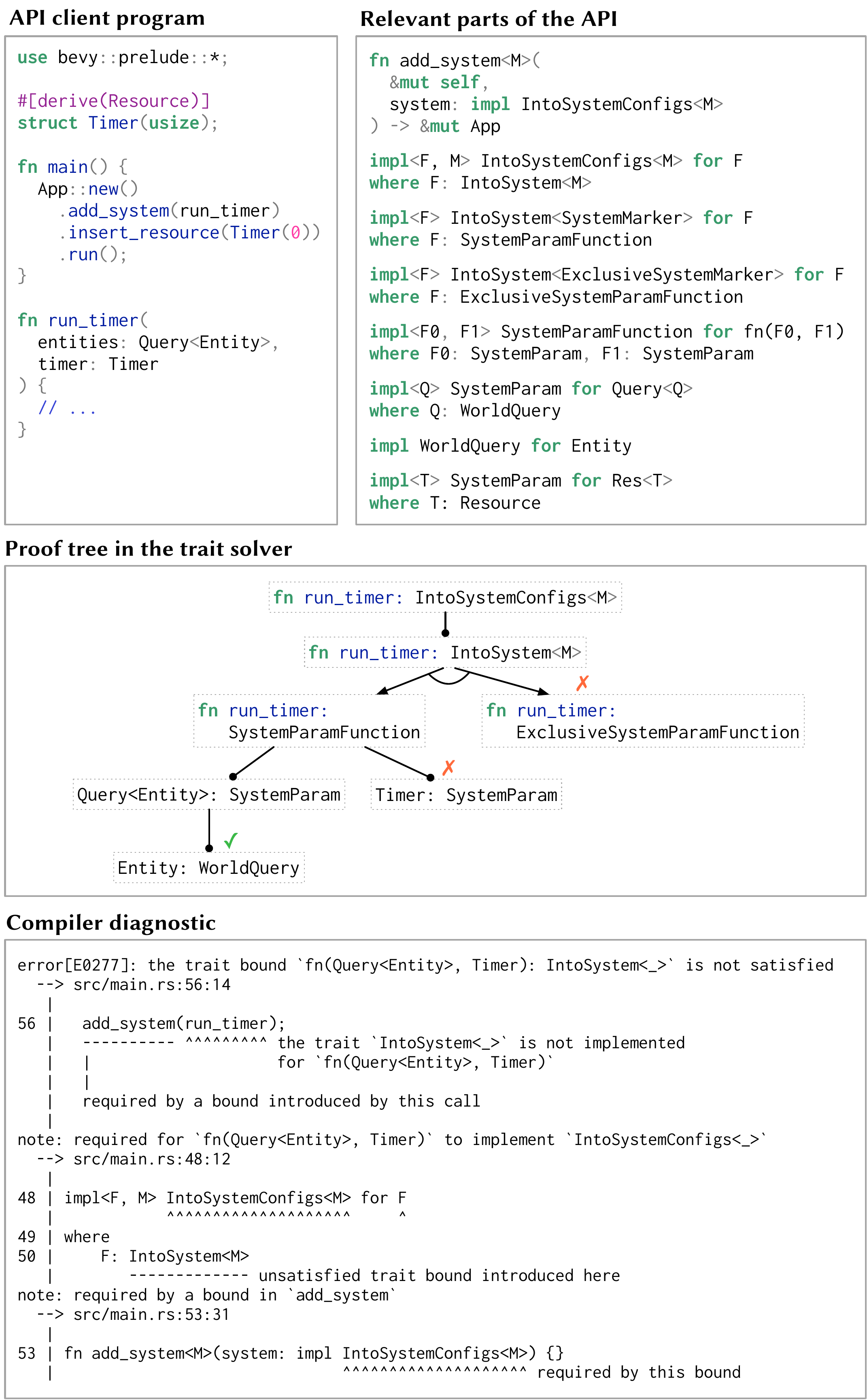}
    \vspace{-0.5em}
    \cprotect\caption{A Rust program (top left) uses the Bevy game engine (top right) with an incorrect type parameter to \rss{run_timer}. The partial proof tree---represented as a feature model diagram---for the \rss{IntoSystemConfigs} trait bound (middle) results in a poor diagnostic (bottom).}
    \label{fig:example-bevy-ptree}
\end{figure}

As a running example, consider the Rust program in \Cref{fig:example-bevy-ptree} (top left). This program uses the Bevy game engine\,\citeurl{https://bevyengine.org/} to create a small game. Bevy uses an entity-component-system (ECS) architecture. The key details relevant to this example are that a \emph{system} is a function which takes as input data provided by the game. For example, \rss{Query<Entity>} provides an iterator over all entities in the world. Bevy also supports \emph{resources}, or global state. To access a resource like \rss{Timer}, a system should take as input \rss{Res<Timer>}, but the \Cref{fig:example-bevy-ptree} program incorrectly takes a plain \rss{Timer}. Based on personal correspondence with the Bevy maintainers, this is a common type of user error.

To check that users provide systems of the correct type, Bevy uses a system of traits shown in \Cref{fig:example-bevy-ptree} (top right). Note that these traits are much simpler than the real Bevy traits, and they omit many implementations of the same traits for irrelevant types. At the top-level, a system must implement \rss{IntoSystemConfigs<M>} for some type \rss{M}. Then \rss{IntoSystemConfigs<M>} is implemented for types implementing \rss{IntoSystem<M>}. Here we branch: types that implement \rss{SystemParamFunction} implement \rss{IntoSystem<SystemMarker>}, and types that implement \rss{ExclusiveSystemParamFunction} implement \rss{IntoSystem<ExclusiveSystemMarker>}. A function \rss{fn(F0, F1)} implements \rss{SystemParamFunction} if both its parameters' type, \rss{F0} and \rss{F1}, implement \rss{SystemParam}. The \rss{SystemParam} trait is implemented for \rss{Res<T>} if \rss{T: Resource} and for \rss{Query<Q>} if \rss{Q: WorldQuery}.

The preceding paragraph contains many details! Again, we stress that this example is still small in comparison to real-world Rust. The program is barely a ``Hello World'' for Bevy, and the traits used simplify many gritty details in Bevy's implementation. Nonetheless, this example is sufficient to demonstrate the diagnostic issue. Within the Rust compiler, the trait solver constructs a proof tree like the diagram in \Cref{fig:example-bevy-ptree} (middle). The diagnostic system consumes this proof tree and then emits the error shown in \Cref{fig:example-bevy-ptree} (bottom). An ideal diagnostic would point to the fact that \rss{Timer} needs to implement \rss{SystemParam}, but this diagnostic instead points farther up the proof tree. It says that the entire function signature does not implement \rss{IntoSystem<_>}, providing no further details.

At a high level, the issue here is the branching point at \rss{IntoSystem<M>}. The diagnostic system does not know whether the user intended \rss{run_timer} to implement \rss{SystemParamFunction} or \rss{ExclusiveSystemParamFunction}, and so the diagnostics cut off at that point in the proof tree. This differs from the \rss{ToString} example where there only exists a single implementation possibility.
More generally, the diagnostic system is full of what we call ``complexity cliffs.'' Error messages contain helpful debugging information at small levels of complexity. Once a complexity threshold is exceeded, then error messages degrade in quality, akin to the infamous Prolog error: \verb|No.|

This trend of poor compiler diagnostics has led to libraries implementing their own debugging tools to get more precise errors. For the Bevy game engine this tool is called bevycheck \citegithub{https://github.com/jakobhellermann/bevycheck}, a simple debugging macro that statically checks that all type requirements of \rss{SystemParamFunction} are met. The requirements themselves are straightforward, every parameter needs to satisfy the trait bound \rss{SystemParam}. This crate-level debugging intervention is not only used in Bevy. Other popular trait-heavy crates, Axum\,\citegithub{https://github.com/tokio-rs/axum}, and Diesel\,\citegithub{https://github.com/diesel-rs/diesel} provide this same macro solution. Moreover, the Rust Foundation has started funding efforts to improve error messages for trait-heavy crates\,\citegithub{https://github.com/weiznich/rust-foundation-community-grant}.

\section{A Trait Debugger}

We are working to address this problem by designing a \emph{trait debugger}. The broad goal is to extract the internal state of the Rust trait solver, and then present it to the developer in the form of a broken proof tree as an aid to debugging. The hypothesis is that the proof tree can provide more granular information about the root cause of an error compared to reporting a failure of the top-level trait bound, and it can do so in a domain-general way. The advantage of proof trees over traditional diagnostics is that they can scale to larger programs and not fail at complexity cliffs. The proof tree shown in \Cref{fig:example-bevy-ptree} (middle) is represented using the notation common to feature model diagrams \cite{KangFeatureOrientedDomain1990}. Goals are shown in the nodes. \textit{Mandatory} subgoals are shown with a small filled circle above them (e.g., all subgoals of \rss{run_timer: SystemParamFunction} must hold). \textit{Alternative} subgoals are shown as children of the same parent goal with an arc drawn through their edges (e.g., one subgoal of \rss{run_timer: IntoSystem<M>} must hold). We have built an initial prototype of this debugger that produces trees similar to that shown in \Cref{fig:example-bevy-ptree} (middle), although the output is currently not quite as polished as the figure. 

A concern from our initial prototype is that proof tree may become too large to effectively skim, reducing their value as a debugging aid. This is more of a statement on the structure of modern Rust crates---Bevy contains thirty-four parameterized implementors for the trait \rss{SystemParamFunction}, all of which are represented in the raw proof tree. Therefore, even after pruning implementation details, the proof tree  will still likely need to be augmented with heuristics that suggest ``starting points'' for exploration. As is the case in the code of \Cref{fig:example-bevy-ptree}, the function \rss{run_timer} was never meant to implement \rss{ExclusiveSystemParamFunction} and the proof tree reflects this. Traditional diagnostics have a hard time moving past these barriers but given the information in the proof tree it's easier to drill down to the root cause of the trait error, the unsatisfied bound \rss{Timer: SystemParam}. Proof trees make finding unsatisfied bounds easier, and the provenance for introduced bounds comes baked in the structure. 

There are several key challenges in developing this tool as proposed. The major challenge is how to compactly and interactively visualize the tree. So far we've described additional diagnostic information as being good, but too much information and developers can feel overwhelmed. Trees with too much information will lead to wasted time searching through the extra nodes and ultimately a difficult-to-use debugger. The last challenge is to maintain the façade that Rust has built, developers should not know they are debugging Prolog programs. Displayed information needs to be reported in terms of the source program, referring to source locations when appropriate. 

In practice we've found the last challenge particularly difficult. The raw proof trees obtained directly from the trait solver reflect implementation details. Fixpoint iterations and performance optimizations are details we need to abstract out of the proof trees before presenting them to the user. How to properly build these abstractions while retaining the proof tree semantics is an area where we are actively working.

We will demonstrate the current state of our debugger at HATRA. We are looking for any feedback regarding design decision, suggestions to visualize the trees, and further connections between this problem and other work.

\section{Related Work}

The challenge of debugging Rust trait errors overlaps with several areas of related work: type inference diagnostics, logic program debuggers, and human factors of proof assistants. One goal of this paper is to pick out ideas from these areas which can influence the design of our trait debugger, as well as to solicit missed connections from readers. 

\subsection{Diagnosing Type Errors}

Hindley-Milner type inference is at once a great triumph of functional programming, and simultaneously a source of unending pain. For 40 years, researchers have proposed increasingly sophisticated methods for diagnosing type inference errors (although none have made it to production, to our knowledge). A variety of strategies have emerged:

\paragraph{Fault localization.}

One strategy is to blame the ``right'' line of code for a type error, which is usually not where the initial error is noticed by the type-checker. \citet{wand1986findingTE} developed an algorithm to track the provenance of unifications made by the type-checker, which could then be presented to the user. \citet{hage2007heuristic} used heuristics such as a ``trust factor'' to sort type constraints in order from most to least problematic. Many recent systems look for sets of constraints such that the program would type-check. \citet{pavlinovic2014type} and \citet{loncaric2016type} use an SMT solver, and \citet{meyers2015class} use a Bayesian analysis. \citet{seidel2017learn} use machine learning to predict blame based on a training set of ill-typed programs.

A fault localization approach could help with diagnosing Rust trait errors. In the \Cref{fig:example-bevy-ptree} example, the root cause was a particular failed inference that could be identified by a heuristic like ``the deepest failed inference in the proof tree.'' However, such a heuristic-based approach is unlikely to generalize to all possible Rust trait errors---after all, that is exactly how the compiler developers have operated for many years, and their large bag of heuristics is not sufficient for cases like \Cref{fig:example-bevy-ptree}. Nonetheless, we expect that some heuristics will be valuable in filtering and ranking the information displayed in our interactive proof tree.

\paragraph{Interactive debuggers.}

Rather than try to find the one true answer, an alternative is to give the programmer an interface into all the information in the type system. \citet{chitil2001composition} designed an ``explanation graph'' with a command-line interface where users ask about the constraints and computed types of sub-expressions in a process called ``algorithmic debugging.''  \citet{asai2013embedded} refine this approach to not require a debugger-friendly reimplementation of type inference. \citet{stuckey2003chameleon} refine this approach by tracking the provenance of constraints as a debugging aid. \citet{chen2014guide} develop a similar ``guided type debugging'' approach that leverages counterfactual typing\,\cite{chen2014conterfactual} to more quickly identify a solution.

We envision our proof tree tool to certainly be interactive. Like these tools, it should be able to explain the proof tree in terms of the source program with a rich mapping between the two. Unlike these tools, we intend to develop a more expressive 2D graphical interface rather than being restricted to text on the command line. 

\paragraph{Automated repair.}

Several systems for type error diagnosis attempt to identify a small change to the input program that causes it to become well-typed\,\cite{lerner2007search,chen2014conterfactual,sakkas2020repair}. This repair can then either directly solve the user's type error, or point them to the root cause. Repair may help with Rust trait errors, but it seems premature to reach directly for program synthesis until we have exhausted the obvious avenues for a diagnostic-only tool.

\paragraph{Domain-specific annotations.} Rather than building a fully generic diagnostic system, an alternative is to give library authors the necessary tooling to express domain-specific knowledge about common trait errors. \citet{heeren2003scripting} describe such an approach for the Helium subset of Haskell where library authors create domain-specific type rules with error messages tailored to the library's domain. Notably, most of the efforts in the Rust ecosystem towards addressing trait errors also have the shape of domain-specific annotations. RFC \#2397\,\citeurl{https://github.com/rust-lang/rfcs/blob/master/text/2397-do-not-recommend.md} describes a \rss{#[do_not_recommend]} annotation that library authors could place on certain trait implementations. For instance, if a trait is implemented for tuples of length 32, then a library author could mark that implementation to not appear in the suggestions of diagnostics. 
A domain-specific approach could likely work in complement to a domain-general approach like we propose. 

\subsection{Logic Programming}

\begin{figure}
    \centering
    \includegraphics[width=0.8\textwidth]{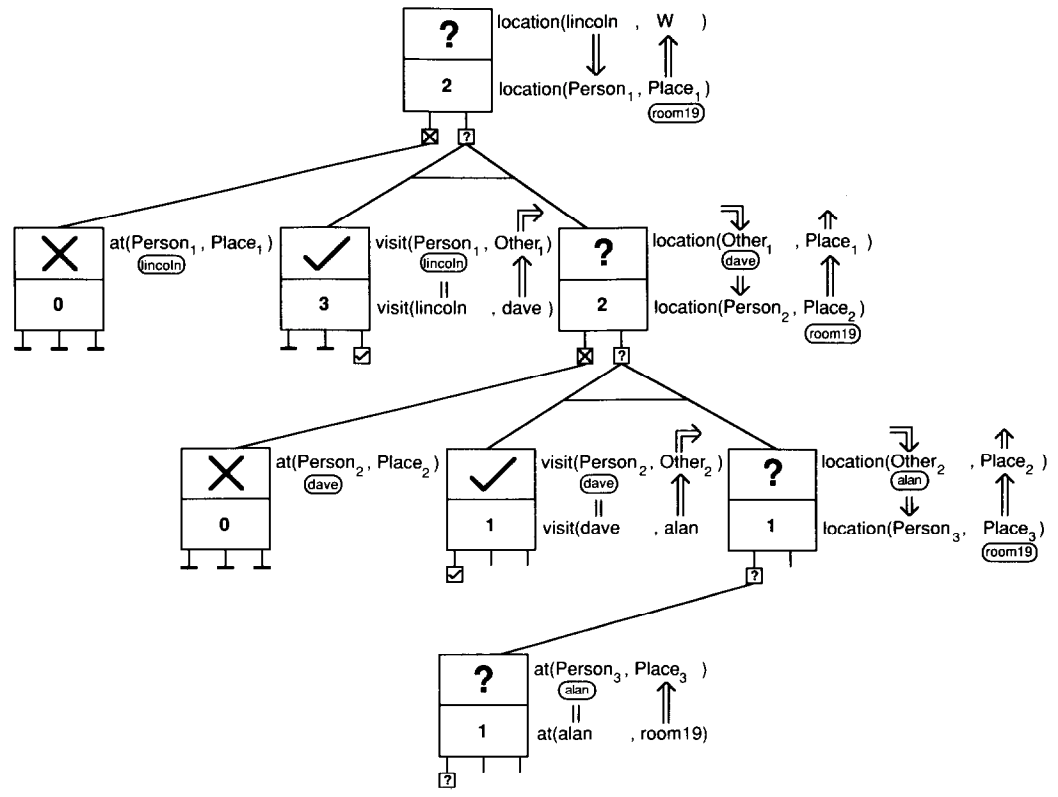}
    \includegraphics[width=0.8\textwidth]{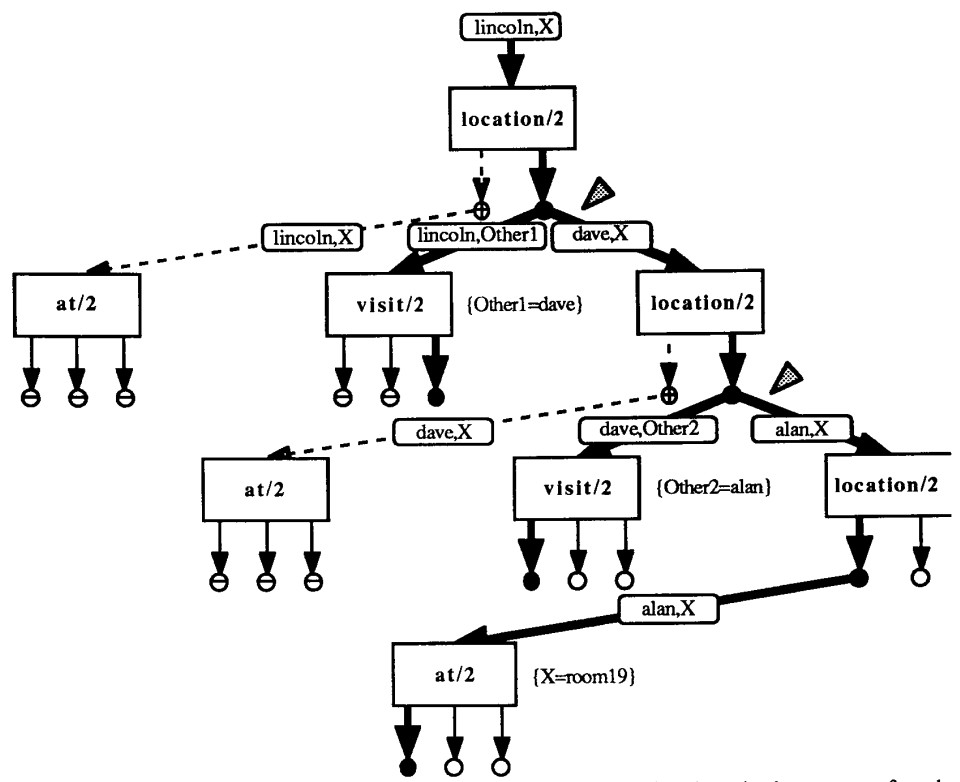}
    \caption{Examples of logic program trace visualizations. Top: an AORTA tree from \citet{eisenstadt1988transparent}. Bottom: a cyclic AND/OR tree from \citet{senay1991logic}.}
    \label{fig:prolog-vis}
\end{figure}

While logic programming has fallen out of fashion, the sizable research program around it in the 1980s and 90s has left us many interesting threads to potentially pick back up. In particular, researchers developed a number of tools to facilitate debugging of Prolog programs. Several systems visualized ``and/or trees'' that represented the execution trace of a Prolog program: the Dewlap debugger\,\cite{dewar1986prolog}, the Transparent Prolog Machine\,\cite{eisenstadt1988transparent}, and cyclic AND/OR graphs\,\cite{senay1991logic}. \Cref{fig:prolog-vis} shows examples of diagrams from the latter two systems. Other Prolog debuggers like Opium\,\cite{ducasse1998opium} focused on abstracting data and control-flow within large execution traces.

These trees provide some guidance in how to compactly visualize various aspects of a logic program trace, such as unification and backtracking. We expect to adopt some of these techniques into our visual design for the trait debugger. However, the challenge will be to ensure the visualization scales to more complex programs, in the sense that a person can read through the diagram and find the information they are looking for. 

\subsection{Debugging Proof Assistants}

The use of tactics in modern proof assistants seems to have the same flavor of usability problem as trait errors. For example, a programmer applies a tactic to some goal, and either succeeds or gets simply a ``No'' from the proof assistant. In theory, similar techniques might be useful to diagnose a trait error as to diagnose a tactic failure. However, we struggled to find much related work on this subject. \citet{shi2023tactic} describe a ``tactic preview'' which can help readers of proofs identify the goals solved by a given tactic. Beyond this, we are interested to find any additional research about tools for debugging broken proof trees.


\bibliography{bib}


\end{document}